\documentclass[preprint]{aastex}

\def\ls{\mathrel{\lower0.6ex\hbox{$\buildrel {\textstyle <}
 \over {\scriptstyle \sim}$}}}

\shorttitle{Radio Source Contribution to Microwave Sky}
\shortauthors{Boughn \& Partridge}

\begin{document}

\title{RMS Radio Source Contributions to the Microwave Sky}

\author{S. P. Boughn and R. B. Partridge}
\affil{Department of Astronomy, Haverford College, Haverford, PA  19041;
sboughn@haverford.edu}

\begin{abstract}
Cross-correlations of the WMAP full sky K, Ka, Q, V, and W band maps with
the 1.4 GHz NVSS source count map and the HEAO I A2 2-10 keV full sky X-ray flux
map are used to constrain {\it rms} fluctuations due to unresolved microwave sources
in the WMAP frequency range.  
In the Q band (40.7 GHz), a lower limit, taking account of only those fluctuations 
correlated with the 1.4 GHz radio source counts and X-ray flux, corresponds to
an {\it rms} Rayleigh-Jeans temperature of $\sim 2 \mu$K for a solid angle of 
one square degree.  The correlated fluctuations at the other bands are consistent with 
a $\beta = -2.1 \pm 0.4$ frequency spectrum.  Using the {\it rms} fluctuations of the 
X-ray flux and radio source counts, and the cross-correlation of these two quantities
as a guide, the above lower limit leads to a plausible estimate of $\sim 5 \mu$K 
for Q-band {\it rms} fluctuations in one square degree.  This value is 
similar to that implied by the excess, small angular 
scale fluctuations observed in the Q band by WMAP, and is consistent with estimates 
made by extrapolating low-frquency source counts.
\end{abstract}


\section{Introduction} \label{intro}
The only available all-sky survey for radio sources at frequencies above 5 GHz is the
WMAP survey \citep{hin07}.  The WMAP team has cataloged several hundred radio sources 
detected at flux levels $> 1$ Jy in one or more of its five frequency bands; a refined 
search by Lopez-Caniego {\it et al.} (2007) has increased the number to nearly 400. 
Virtually all of these sources are already known from lower frequency radio catalogs
such as FIRST, GB6, and the PMN survey in the southern hemisphere (see, for example,
Trushkin 2003, Hinshaw {\it et al.} 2007, and Lopez-Caniego {\it et al.} 2007).  
The WMAP sources are a mix of flat spectrum sources, 
such as quasars, and the brightest ``classical'' radio galaxies that may have steep
spectra, but retain adequate flux in one or more of the WMAP frequency bands.

Since the WMAP sources do appear in lower frequency catalogs with high probability, 
it follows that correlating lower frequency catalogs with the WMAP images can either
identify additional sources, as was found by Lopez-Caniego {\it et al.} (2007), or
can reveal the statistical properties of sources too weak to be individually detected
in the WMAP images.  In particular, a cross-correlation of the WMAP images with lower
frequency catalogs can provide information on {\it rms} fluctuations induced by such 
sources at each of the WMAP frquencies, and thus average spectral indices for these
subliminal sources can be deduced.  Since the subliminal WMAP sources that dominate 
the fluctuations lie roughly in the decade $0.1-1$ Jy, they are quite likely to 
appear in lower frequency catalogs, which generally reach to much fainter flux 
densities.  This paper treats the correlation of WMAP images
at 23, 33, 41, 61, and 93 GHz with number counts from the NVSS 1.4 GHz survey \citep{con98}.
Since many flat spectrum radio sources are known to be X-ray emitters, we also
cross-correlate the WMAP images with the HEAO I A2 2-10 keV all-sky X-ray survey.

\section{Data Sets} \label{data}
\subsection{HEAO and NVSS Maps}
The versions of the HEAO and NVSS maps used in the present analysis were used previously in a 
variety of analyses including: an investigation of the large-scale structure of the X-ray background 
\citep{bou02a}; a determination of the large-scale bias of the X-ray background \citep{bou04a};
and measurements of the ISW effect \citep{bou02b,bou04b}.

The NRAO VLA Sky Survey (NVSS) is a flux limited survey (nominally at 2.5 mJy) at a frequency of
1.4 GHz \citep{con98}.  It is complete for declinations above $-40^{\circ}$ and 
contains $1.8 \times  10^6$ sources, with a mean source number density of 51.8 per 
square degree.   The source counts were binned in 24,576  $1.3^\circ \times 1.3^\circ$ pixels.  
The pixelization we use is a quadrilateralized spherical cube projection \citep{whi92}.
Those pixels that were only partially contained in the survey region 
were omitted. There are systematic number count offsets that occur in several 
declination bands that coincide with discontinuous changes in the {\it rms} noise levels in
the survey \citep{bou02b}.  To correct for this, random sources were added to or
subtracted from each pixel to eliminate the band structure. The resulting map shows 
no declination dependent structure at a level of $< 1\%$. For comparison, the Poisson
noise per pixel is  11\%.  In order to exclude Galactic sources as well as nearby
clusters of galaxies, the map was masked with the most agressive mask (Kp0) used by the WMAP 
team.  In addition, regions that contained pixels with source counts greater than 
$4 \sigma$ (corresponding to $\ge 43\%$ of the mean number of sources per pixel) 
above the average count of 87.5 were masked.   This procedure cleaned the map of 10 
"objects" located more than $10^{\circ}$ from the Galactic plane. Among these are the 
Orion Nebula and the nearby Virgo, Perseus, and Fornax clusters. That the moderately 
nearby, rich Coma cluster of galaxies was not one of the regions cut out indicates to us that 
this masking removed only Galactic and nearby extra-galactic sources.  As is the case with
the HEAO source masking, the effect of this extra cleaning was primarily to reduce the
noise and did not significantly change the value of the correlation with the WMAP data.
There is a small residual dipole in the NVSS map that is much too large to be that due to
the Earth's motion with respect to the comoving frame of the radio sources and, therefore,
is undoubtedly due to some large scale systematic.  This dipole moment was fit and 
removed from the map.

The HEAO1 A2 data set \citep{bol87} we employ here was constructed from the output of two 
medium energy detectors (MED) with different fields of view ($3^\circ \times 3^\circ$ and 
$3^\circ \times 1.5^\circ$) and two high energy detectors (HED3) with
these same fields of view.  These data were collected during the six month period
beginning on day 322 of 1977.  Counts from the four detectors were combined, then pixelized
with the same projection used for the NVSS source counts.
The combined map has a spectral bandpass (corresponding to quantum efficiency $> 50\%$) of
approximately $3-17~keV$ \citep{jah89}, but the counts have been
converted to equivalent flux in the more standard $2-10~keV$ band.  Because of the 
ecliptic longitude scan pattern of the HEAO satellite, sky
coverage and therefore photon shot noise is not uniform.  However, the
variance of the cleaned, corrected map, $2.1 \times 10^{-2}$(TOT cts s$^{-1})^2$,
is significantly larger than the variance of photon shot noise,
$0.8 \times 10^{-2}$(TOT cts s$^{-1})^2$, where
1 TOT cts s$^{-1} \approx$ 2.1 $\times 10^{-11}$ erg s$^{-1}$ cm$^{-2}$ 
\citep{all94}.  This implies that most of the variance in
the X-ray map is due to ``real'' structure.  For this reason, and
to reduce contamination from any systematics that might be correlated with the scan
pattern, we chose to weight the pixels equally.

As with the NVSS map, the HEAO map was masked with the WMAP Kp0 mask and then more
aggressively masked by 
removing all pixels within $20^{\circ}$ of the Galactic plane and within 
$30^{\circ}$ of the Galactic center.  In addition, large regions 
($6.5^\circ \times 6.5^\circ$) centered on $92$ nearby, discrete X-ray sources 
with $2 - 10~keV$ fluxes larger than $3 \times 10^{-11}$ erg s$^{-1}$ cm$^{-2}$ 
\citep{pic82} were removed from the maps.  Around the sixteen brightest of 
these sources (with fluxes larger than $1 \times 10^{-10}$ erg s$^{-1}$ cm$^{-2}$),
the masked regions were enlarged to $9^\circ \times 9^\circ$.   We also used the
ROSAT All-Sky Survey (RASS) Bright Source Catalog \citep{vog96}
to identify additional bright 0.5-2 keV sources.  While the RASS survey has somewhat
less than full sky coverage ($92\%$), it has a relatively low flux limit that 
corresponds to a $2-10~keV$ flux of  $\sim 2 \times 10^{-13}$ erg s$^{-1}$ cm$^{-2}$ 
for a photon spectral index of $\alpha = -2$.  Every source in the RASS catalog was
assigned a $2-10~keV$ flux from its hard channel (0.5-2.0 keV) flux by assuming a spectral 
index of $-3< \alpha < -1$ as deduced from the hardness ratio within this band.   There
were 34 sources with fluxes exceeding  $3 \times 10^{-11}$ erg s$^{-1}$ cm$^{-2}$
that were not in the Piccinotti source list, and
these were also masked.  The HEAO I A2 map itself was searched for ``sources'' that 
exceeded the nearby background by a specified amount.  Thirty additional such
sources at a level of $\sim 3 \times 10^{-11}$ erg s$^{-1}$ cm$^{-2}$ were identified 
and masked.  These final two cuts amount to only 6.5\% of the sky and, in any case,
do not significantly change the values of the correlations presented in \S\ref{corr}.  The primary
reason for masking nearby, strong sources is to reduce the noise in the correlations
discussed in \S\ref{corr}.  The final
masked map has 48\% sky converage.  Finally, several systematic foreground features
were fit and removed from the map: a linear time drift of detector
sensitivity, high latitude Galactic emission, the dipole induced by the earth's 
motion with respect to the X-ray background, and emission from the plane of the local 
supercluster.  These corrections are discussed in detail in Boughn (1999) and
Boughn {\it et al.} (2002).

\subsection{Microwave Maps}
The microwave data consists of the five frquency bands, K (centered at 22.7 GHz), Ka (33.0 GHz), 
Q (40.7 GHz), V (60.6 GHz), and W (93.4 GHz), of the three year WMAP intensity maps 
\citep{hin07}.  These maps were also converted from the available Res 9 HEALPix
format to the same $1.3^{\circ}  \times 1.3^{\circ}$ quadrilateralized spherical cube 
pixelization of the other two maps.  In order to exclude the strong microwave emission from
the Galactic plane, all maps were masked with the agressive WMAP Kp0 mask.  
We note that, in addition, this mask
removes all extragalactic microwave sources with fluxes greater than 1 Jy.  For comparison,
we also analyzed the maps without masking these sources, as discussed in \S\ref{sys}. 
All five maps were corrected for the cosmic microwave background emission, which would
otherwise add noise to the cross correlations, by subtracting the WMAP ILC CMB map. 
While this correction reduces the noise slightly, it does not have a significant effect
on the lower limits deduced in \S\ref{con}.  
The corrected maps still have a significant component of high latitude Galactic
emission, especially on large angular scales.  Since fluctuations due to extragalactic 
sources are predominently on small angular scales, the maps were high passed filtered by 
simultaneously fitting and removing the dipole, quadrupole, and octopole moments.  A Galactic 
secant law and the $H\alpha$ map compiled by Finkbeiner (2003), a template for Galactic 
free-free emssion, were included in the fits in order to remove additional Galactic emission.
The resulting masked maps were then smoothed with a $24^{\circ}$ FWHM Gaussian filter ignoring
the contributions of masked pixels.  These smoothed 
maps were subtracted from the corrected maps in order to provide additional high pass
filtering.   For unmasked pixels near large masked regions, the high pass filtering
is somewhat reduced; however, only 1\% of these pixels have the effective smoothing area 
reduced to a value below 0.4 times the nominal and no pixels had effective smoothing areas 
reduced below 0.11.  In
any case, the effects of this high pass filter on small scale fluctuations were minimal, 
as discussed in \S\ref{sys}.

\section{Correlations} \label{corr}
Since the primary objective of the present analysis is to constrain the fluctuations of the 
equivalent brightness temperature due to microwave sources, it might seem reasonable to simply compute
the {\it rms} fluctuations of the various WMAP maps.  However, even with the masking, filtering, 
and corrections discussed in \S\ref{data}, the {\it rms} fluctuations are completely dominated by 
Galactic emission.  In order to isolate that component due to distant extragalactic sources,
we cross-correlate the microwave maps with our two templates of extragalactic sources, the NVSS 
radio source count map and the HEAO X-ray flux map, which is dominated by distant AGN.  The redshift
distribution of the signals represented by both of these maps are broad with a peak redshift 
of $z \sim 1$ \citep{con84,dun90,bou04a}.  A standard measure of the correlation of two data 
sets is the cross-correlation function (CCF), which in this case is defined by
\begin{equation}
CCF(\theta) = {1 \over N_{\theta}} \sum_{i,j} (S_i -\bar{S})(T_j -\bar{T})
\label{eq:ccf}
\end{equation}
where the sum is over all pairs of pixels $i,j$ separated by an angle 
$\theta$, $S_i$ is the number of NVSS sources (or the X-ray intensity) in the 
$i$th pixel, $\bar{S}$ is the mean number of sources (X-ray intensity), $T_i$ is the WMAP 
equivalent thermodynamic temperature of the $i$th pixel, $\bar{T}$ is the mean 
temperature, and $N_{\theta}$ is the number of pairs of pixels separated by $\theta$.
In order to interpret the cross-correlations of the two templates with the microwave 
maps, it is necessary to determine the {\it rms} fluctuations in the template maps as well
as the cross-correlation of these two maps.  The auto-correlation function (ACF) is defined,
using the same notation as in Equation \ref{eq:ccf}, as
\begin{equation}
ACF(\theta) = {1 \over N_{\theta}} \sum_{i,j} (S_i -\bar{S})(S_j -\bar{S}) .
\label{eq:acf}
\end{equation}
Figure 1 and Figure 2 depict the ACFs of the NVSS source counts and
X-ray flux.
\begin{figure}
\begin{center}
\includegraphics[scale=1.0]{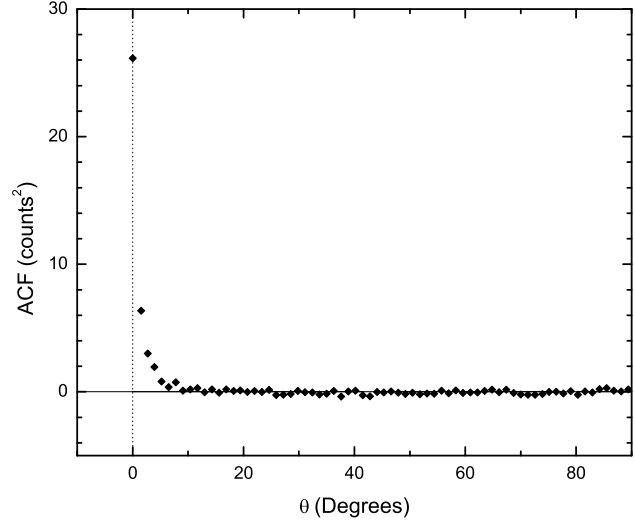}
\caption{The auto-correlation function of the NVSS map corrected
as discussed in the text.  The units are (counts/pixel)$^2$.}
\end{center}
\label{fig:nacf}
\end{figure}
\begin{figure}
\begin{center}
\includegraphics[scale=1.0]{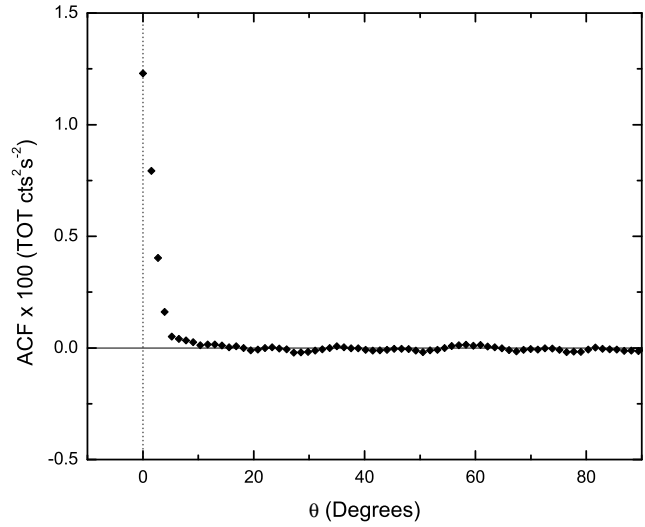}
\caption{The auto-correlation function of the HEAO 1 A2 2-10 keV X-ray map corrected
as discussed in the text.  1 TOT cts s$^{-1} = 3 \times 10^{-11}$ erg s$^{-1}$ cm$^{-2}$.}
\end{center}
\label{fig:xacf}
\end{figure}
We have corrected the ACF(0) point of Figure 1 for Poisson noise by
subtracting 87.5, the mean number of sources per pixel.  With this correction, the signal
at all angles is an indication of the angular clustering of NVSS radio sources 
\citep{bou02b}.  The X-ray ACF at zero lag has been corrected for photon shot noise, but not for the 
Poisson noise in the distribution of sources since the resolution of the map, 3.04 degrees, 
is significantly larger than a pixel.  However, a model of X-ray clustering indicates
that the fraction of ACF(0) due to Poission noise is approximately 77\%, coincidentally
the same as for the NVSS ACF(0).  Figure 3 is the NVSS/HEAO CCF.  The 
(highly correlated) errors in these ACFs and the CCF have not been plotted; however, the signal 
to noise values are large, i.e., S/N $> 10$ for $\theta \le 3^{\circ}$ \citep{bou02a,bou04a}, 
and have an insignificant effect on the analysis of this paper.
\begin{figure}
\begin{center}
\includegraphics[scale=1.0]{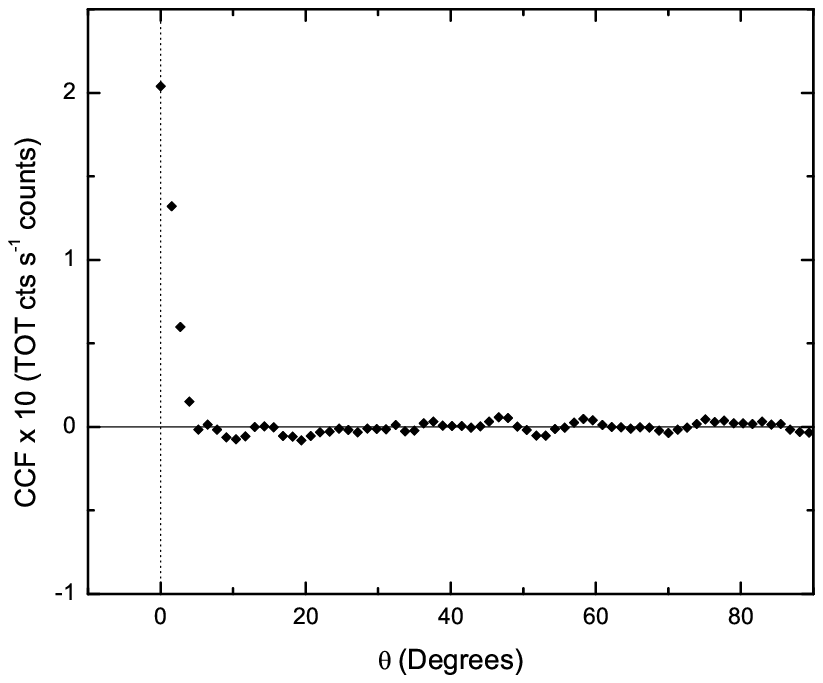}
\caption{The cross-correlation function of the X-ray and NVSS maps of Figures 1 
and 2}
\end{center}
\label{fig:xnccf}
\end{figure}

Figure 4 is the CCF of the corrected NVSS map with the WMAP Q-band map.  It is 
clear that only the small angular scale fluctuations of the two maps are correlated.  Since the
noise in the WMAP maps is not well characterized at the level of interest in our analysis, 
$\sim 1 \mu$K, we chose to estimate the noise by computing 240 CCFs of maps that were rotated
with respect to each other.  Five rotations were performed about each of 48 uniformly spaced axes 
separated by $\sim 21$ degrees.  The resulting distribution of CCFs is,
to a reasonable approximation, Gaussian  and so the error bars that are plotted in 
Figure 4 have the usual $1~\sigma$ interpretation.  The errors so determined are accurate 
at the 10\% level, which is sufficient for the arguments in \S\ref{con}. 
As is clear from the figure, these errors are significantly correlated.
This CCF is typical of the the CCFs at the other WMAP frequencies.  
\begin{figure}
\begin{center}
\includegraphics[scale=1.0]{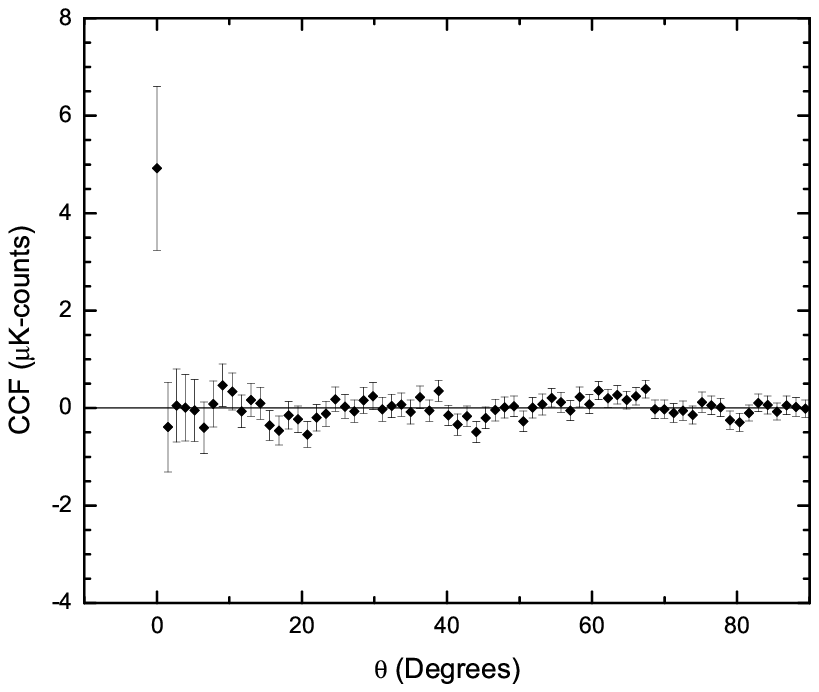}
\caption{The cross-correlation function of WMAP Q band map with the NVSS map.  Errors are determined
from 240 rotated versions of the data and are highly correlated.}
\end{center}
\label{fig:qnccf}
\end{figure}
Figure 5 shows
the dependence of CCF(0) on frequency.  The curve represents a power law microwave spectrum
typical of free-free emission, $T \propto \nu^{-2.1}$, and it is clear that the data are consistent 
with such a model, although there is no clear detection at the highest (W-band) frequency.
The curve is not a fit to the data but is forced to go through the highest signal to noise 
point at 33 GHz.  Formal fits to these data are discussed in \S\ref{con}.
Figure 6 and Figure 7 are the corresponding cross-correlation 
between Q-band and X-ray maps, and the CCF at zero lag for all WMAP frquencies.  
It is clear that the detection of these correlations
is $2 \sigma$ at best for the K, Ka, and Q bands, with no significant detection at the  V and W bands.
\begin{figure}
\begin{center}
\includegraphics[scale=1.0]{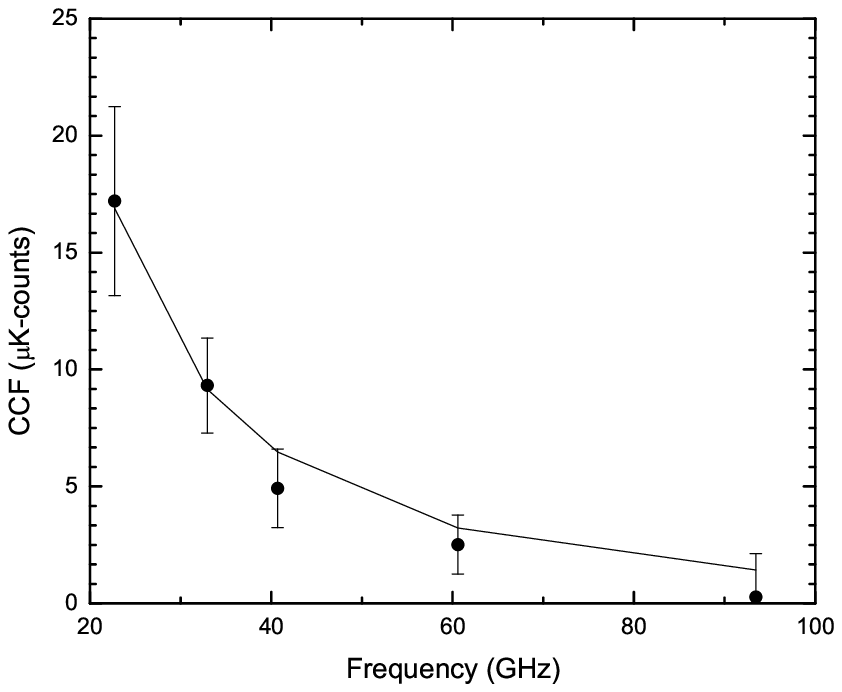}
\caption{CCF at zero lag of all 5 WMAP maps with the NVSS map.}
\end{center}
\label{fig:nalccf}
\end{figure}
\begin{figure}
\begin{center}
\includegraphics[scale=1.0]{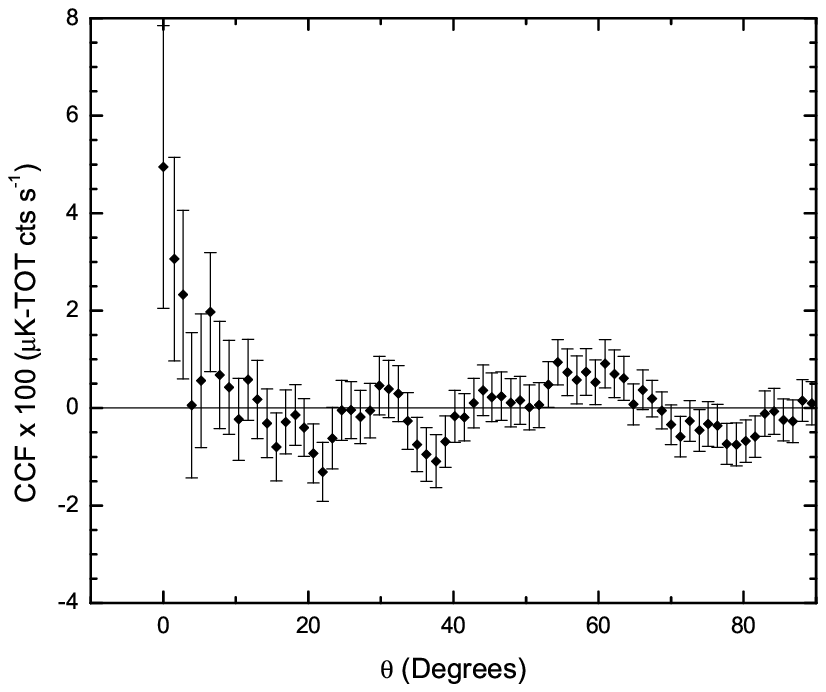}
\caption{The cross-correlation function of WMAP Q band map with the X-ray map.  Errors are determined
from 240 rotated versions of the data and are highly correlated.}
\end{center}
\label{fig:qxccf}
\end{figure}
\begin{figure}
\begin{center}
\includegraphics[scale=1.0]{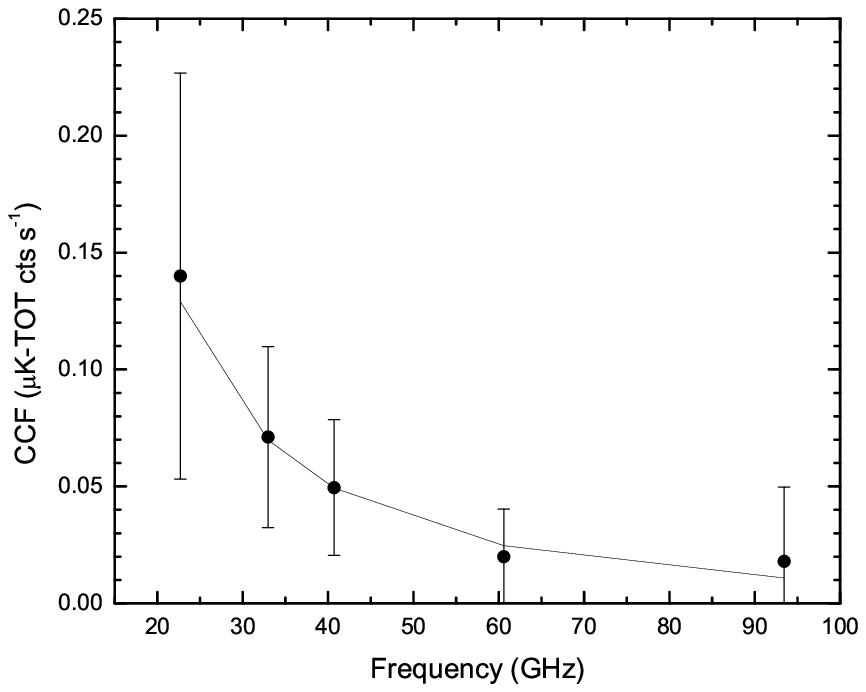}
\caption{CCF at zero lag of all 5 WMAP maps with the X-ray map.}
\end{center}
\label{fig:xalccf}
\end{figure}

\section{Constraints on Microwave Fluctuations} \label{con}
In order to constrain fluctuations in the microwave sky we assume that the microwave emssion $M$ in a
given pixel consists of a signal $aN$ proportional to the number $N$ of NVSS sources in the pixels plus 
a signal $C$ that is statistically independent of $N$, i.e., $M = aN + C$.  Then the zero lag value 
of the cross-correlation of $M$ with $N$ is given by 
\begin{equation}
CCF(0) = \langle MN \rangle = a\langle N^2 \rangle
\end{equation}
where $\langle \rangle$ indicates an ensemble average, in our case to be approximated by the average 
over the maps.  $\langle N^2 \rangle$ is just the NVSS auto-correlation function, which we have measured.  
Solving for for the unknown constant of proportionality gives 
$a = \langle MN \rangle / \langle N^2 \rangle$.  
Finally the mean square microwave fluctuation is
\begin{equation}
\langle MM \rangle = a^2\langle N^2 \rangle + \langle C^2 \rangle + 2a\langle NC \rangle .
\label{eq:nvar}
\end{equation}
Substituting for the constant $a$ and noting that, by assumption, $\langle NC \rangle = 0$, 
the {\it rms} microwave fluctuations are given by 
\begin{equation}
\sigma_M = \sqrt{\langle MN \rangle ^2/ \langle N^2 \rangle + \langle C^2 \rangle} >
\langle MN \rangle/\sqrt{\langle N^2 \rangle} .
\label{eq:nsig}
\end{equation}
Therefore, a lower limit to the microwave fluctuations is given by the WMAP/NVSS CCF(0) divided by the
square root of the NVSS ACF(0).  The error in this lower limit is just the error in the CCF divided by
the same factor, $\sqrt{\langle N^2 \rangle}$.  Table 1 lists these lower limits for all 
the WMAP frequencies along with those deduced in similar fashion from the WMAP/X-ray cross-correlation.  
No attempt is made to interpret the values of
the CCF($\theta$) for $\theta \ne 0$ since there is no detectable signal in the NVSS CCFs and the finite
angular resolution (3.04 degrees FWHM) of the X-ray maps is responsible for most of the 
signal at $\theta > 0$.
\begin{deluxetable}{cccccc}
\tablecaption{Lower Limits on Micrwave Fluctuations Deduced from NVSS and HEAO 
Cross-correlations.  The units are $\mu$K of {\it rms} Rayleigh-Jeans temperature. \label{tbl-1}}
\tablewidth{0pt}
\tablehead{
\colhead{WMAP Band} & \colhead{Secondary Map} & \colhead{Lower Limits}
}
\startdata
K Band (22.7 GHz) & NVSS & 3.31 $\pm$ 0.78  \\
  & HEAO & 2.57 $\pm$ 1.61  \\
  & Combined & 3.24 $\pm$ 1.06  \\
Ka Band (32.95 GHz) & NVSS & 1.77 $\pm$ 0.39  \\
  & HEAO & 1.29 $\pm$ 0.71  \\
  & Combined & 1.77 $\pm$ 0.41  \\
Q Band (40.70 GHz) & NVSS & 0.92 $\pm$ 0.31  \\
  & HEAO & 0.88 $\pm$ 0.52  \\
  & Combined & 0.85 +0.45 -0.26  \\
V Band (60.60 GHz) & NVSS & 0.45 $\pm$ 0.22  \\
  & HEAO & 0.34 $\pm$ 0.35  \\
  & Combined & 0.40 +0.44 -0.15  \\
W Band (93.44 GHz) & NVSS & 0.04 $\pm$ 0.29  \\
  & HEAO & 0.27 $\pm$ 0.48  \\
  & Combined & 0.24 +0.63 -0.22  \\
\enddata
\end{deluxetable}
It should be noted that the ACF(0) used to deduce the values in Table 1 are those corrected for Poisson 
fluctutations for both the NVSS and X-ray data.  It seems reasonable to do this since it is far more 
likely that the simple model of proportional emission applies to sources that cluster with each other 
than the assumption that the Poisson fluctuations of the two populations are correlated.  Had we
not corrected the ACF(0)s for Poisson fluctuations, then the lower limits in Table 1 would decrease by a 
factor of 2.  The consequences of these alternatives will be discussed in \S\ref{dis}.

While the lower limits derived from the X-ray data are quite noisy, it is interesting that they are 
fortuitously similar to the values derived from the NVSS data.  It is also of interest to estimate the 
lower limit of microwave fluctuations by considering the microwave emission that is associated with 
{\it either} NVSS {\it or} X-ray sources.  It's straightforward to expand our simple model by expressing the microwave
emission $M$ as a combination of emssion proportional to $N$, the number of NVSS source counts, a component
proportional to X-ray flux $X$, and a component uncorrelated with either $N$ or $X$, i.e.,
$M = aN + bX +C$.  Then the two CCFs are $\langle MN \rangle = a\langle N^2 \rangle + b\langle NX \rangle$
and $\langle MX \rangle = a\langle NX \rangle + b\langle X^2 \rangle$.  Since we measure all quantities
in these two expression except $a$ and $b$, it is straightforward to solve for them and then evaluate
the {\it rms} microwave fluctuations as in Equation \ref{eq:nsig}, i.e.,
\begin{equation}
{\sigma_M}^2 = \frac{\langle MN \rangle^2 \langle X^2 \rangle + \langle MX \rangle \langle N^2 \rangle
- 2\langle MN \rangle \langle MX \rangle \langle NX \rangle}{\langle N^2 \rangle \langle X^2 \rangle -
\langle NX \rangle} + \langle C^2 \rangle .
\label{eq:nxsig}
\end{equation}
This combination does not constitute one that in any sense maximizes the signal to noise; rather, the purpose 
is to arrive at an estimate of the {\it rms} microwave fluctuations that are correlated with either the NVSS number
counts or the HEAO X-ray emission.  The lower limits to this estimate of the microwave fluctuations, ignoring
the uncorrelated $\langle C^2 \rangle$ term, are also listed in Table 1.  Because the expression in Equation
\ref{eq:nxsig} is non-linear, the errors are non-Gaussian, especially for the lower signal to noise
entries.  They were determined from the 68\% confidence limits deduced from the 240 CCFs determined from
rotated maps.  That the lower limits implied by the combined data are in agreement with the those determined
separately from the NVSS and X-ray data is consistent with the relatively high degree of correlation of these two 
data sets.  It should be noted that the NVSS and HEAO maps were not masked in the same way, so if there is any
dependence of the ACFs and CCFs on masking then the interpretation of their combination in Equation 6 will be
complicated.  We repeated the analysis including only those pixels present in both the NVSS and HEAO maps. 
The values in Table 1 changed by only a small fraction of the error ($\sim 1/3 \sigma$) with the errors
correspondingly larger because of the smaller number of pixels.

Figure 8 shows the frequency dependence of the lower limit to microwave fluctuations deduced 
from the NVSS/WMAP CCF(0).  The fluctuations have been expressed as equivalent Rayleigh-Jeans temperature
in order to facilitate the comparsion with a power spectrum model.  Because the corrected WMAP maps are still
dominated by Galactic emission, these maps are highly correlated with each other.  The result is that the 
errors depicted in Figure 8 are also significantly correlated.  For example, the correlation coefficient  
of the K and Ka band errors is 0.79, 
of the K and Q band errors is 0.61, and of the Ka and Q band errors is 0.85.  The solid curve is a minimum
$\chi^2$ fit to a free-free ($\beta = -2.1$) power spectrum and the dashed line is a fit to a typical synchrotron 
($\beta = -2.8$) spectrum.  The reduced $\chi^2$ of the former fit is 3.3/4 and 8.6/4 for the latter.  A
free-free spectrum is certainly consistent with the data while a synchrotron spectrum is marginally 
inconsistent with the data. (The reason that the synchrotron fit is less consistent than might appear
from Figure 8 is because of the highly correlated errors.)  One might expect that the model curves
would be straight lines; however, due to the differences in the WMAP beam patterns, each data point 
corresponds to a slightly different smoothing that results in slightly different fluctuations than if all
beams were the same angular scale.  If the spectral index is included
in the fit, then the minimum $\chi^2$ fit yields $\beta = -2.11 \pm 0.41$ with a reduced $\chi^2$ of 3.25/3.
The amplitude of the solid curve in Figure 8 is $1.2(\nu/40.7$ GHz$)^{-2.11}\mu$K.  This value is
well below the level at which systematic effects in the WMAP are characterized, which is the primary 
reason for determining the errors from the data themselves rather than trying to model the noise of the maps. 
Another reason is that the NVSS map has significant systematics of its own, as discussed in \S\ref{data}.
\begin{figure}
\begin{center}
\includegraphics[scale=1.0]{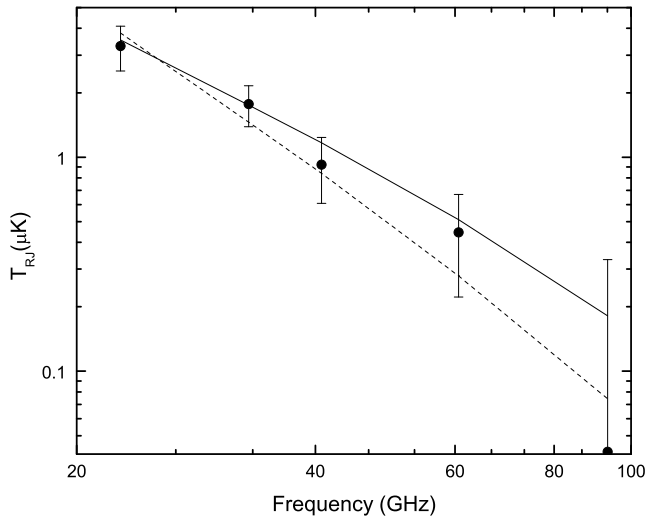}
\caption{Lower limits of the {\it rms} microwave flux in units of Rayleigh-Jeans temperature for
all 5 WMAP frequencies.  The solid curve is the minimum $\chi^2$ fit to the data of a 
$\beta = -2.1$ free-free spectrum and the dashed line is the fit of a $\beta = -2.8$ 
synchrotron spectrum. (See \S\ref{con} for an explanation of the slight curvature of both lines.)}
\end{center}
\label{fig:nlim}
\end{figure}

It should be noted that the lower limits deduced above assumed that the correlations were due entirely to
the joint clustering of NVSS and microwave sources.  If instead we assume that the correlations are due
to the fact that the NVSS and microwave sources are {\it the same} sources, then it would be appropriate 
to substitute the
total, i.e., not Poisson corrected, ACF(0) in Equation \ref{eq:nxsig}.  In this case the the lower limits
are all reduced by a factor of 2.1.  The consequences of this will be discussed in the next section.  
In order to facilitate comparisons with other observations and models we express the above results as the
{\it rms} temperature fluctuations in a one square degree top hat beam.  This value depends
somewhat on the relative contributions of clustering and Poisson noise to the fluctuations.  For example,
if the Poisson process dominates, then the implied lower limit in one square degree is 
$1.9(\nu/40.7$ GHz$)^{-2.1}\mu$K while if the fluctuations are dominated by $1/\theta$ clustering, then the 
lower limit becomes $1.4(\nu/40.7$ GHz$)^{-2.1}\mu$K.  If we take as a more reasonable assumption that 
$\sim 25\%$ of the fluctuations are due to the clustering of sources, comparable to the cases of 
both NVSS and X-ray fluctuations, then the limit becomes  $\sim 1.8(\nu/40.7$ GHz$)^{-2.1}\mu$K.

\section{Discussion} \label{dis}

The results presented in \S\ref{con} are definitive measurements of the fluctuations in the microwave sky
that are correlated either with counts of NVSS radio sources or with HEAO X-ray flux.  Even so, they
are only lower limits on the total fluctuations in the microwave foreground and the values were deduced 
under the assumption that correlations were due to joint clustering of microwave, NVSS, and X-ray sources.
The power law fit to the lower limits indicates that the spectral index is more consistent with free-free
or other flat spectrum than with synchrotron emission, which shouldn't be surprising since high 
frequency observations will naturally 
be biased toward flatter spectrum sources.

\subsection{Estimate of Total Fluctuations}

It is possible to make a (slightly) educated guess as to the total level of microwave fluctuations by 
appealing to the example of NVSS/X-ray correlated fluctuations.  The level of X-ray fluctuations correlated
with NVSS fluctuations can be deduced in the same manner as in Equation \ref{eq:nsig}, i.e., 
$\langle X^2 \rangle > {\langle XN \rangle}^2 / \langle N^2 \rangle$.  The value of this quantity is 
$1.6 \times 10^{-3}$ whereas the total X-ray fluctuation (see Figure 2) is
$ \langle X^2 \rangle = 0.0123$.  Hence $\sqrt{\langle X^2 \rangle}$ is a 
factor of 2.8 larger than the derived lower limit. The same factor applies to the case of a lower 
limit of NVSS fluctuations derived from X-ray fluctuations because Poisson fluctuations happen to account
for the same fraction of total fluctuations in both the NVSS and X-ray cases.  
As was pointed out in \S\ref{data}, the NVSS and X-ray fluctuations have similar redshift distributions.  
\emph{IF} microwave sources also have a similar redshift distribution, \emph{IF}
microwave fluctuations have similar ratios of Poisson to clustering components, and \emph{IF} 
the uncorrelated component ({\it C} in Equation \ref{eq:nvar}) comprises the same fraction of the
X-ray fluctuations as of the microwave fluctuations, then one would expect the same 2.8 correction factor to apply.  
This would imply that the total microwave fluctuations would be $\sim 5~(\nu/40.7$ GHz)$^{-2.1} \mu$K.  
While plausible, there is no direct evidence that the three above conditions are true.  Therefore, this 
conclusion should be considered only as a plausibilty argument.

It was pointed out above that had total ACF(0)'s, including the Poisson contributions, been used to
determine correlated microwave fluctutations, then the derived lower limits would be a factor of 2.1 less than
those listed in Table 1.  A similar analysis using the NVSS/X-ray correlation to determine a lower
limit to the X-ray fluctuations would have have resulted in a lower value by this same factor.  However,
the factor by which the actual X-ray fluctuations exceed the correlated component would have increased
by precisely the same factor.  Therefore, while the lower limits depend on whether the correlated
fluctuations are clustering or Poisson dominated, the rough estimate of {\it total} fluctuations does not.

\subsection{Potential Sources of Systematic Error} \label{sys}

All of the analysis described above was for WMAP maps that were masked for all regions that contained microwave
sources that exceeded 1 Jy.  We repeated the analysis not excluding these sources (but
retaining the mask of the Galactic plane).
The K and Ka band lower limits both increase by a factor of 1.4 while the Q band lower limit increases by a factor
of 1.6.  Given the limited signal to noise, these factor are consistent with each other, and our
conclusions about the spectral index of fluctuations in the microwave sky remain unchanged.

There may be some question as to whether the the high pass filtering discussed in \S\ref{data} above
reduces the CCF(0) and hence the lower limits derived in this paper.  As an estimate of the significance
of this effect, the same procedure described in \S\ref{data} was 
performed on the NVSS map, which was then cross-correlated with 
the unfiltered NVSS map.  The result was a negligible (1.5\%) decrease from the ACF(0) value of the
original NVSS map.  This should be an indication of the level of attenuation expected from the high 
pass filtered WMAP cross-correlated with the NVSS map.  The high pass filtered NVSS map was also 
cross-correlated with the HEAO map in order to estimate the attenuation expected from the
high pass filtered WMAP cross-correlated with the HEAO map.  This attenuation was $\sim 2\%$, again negligible.  
Since the resolution of the K and Ka bands (0.88 and 0.66 degrees) are not negligible compared to the pixel size
(1.3 degrees), the former test might not be appropriate for the filtered  WMAP/NVSS CCF.  However, when a 
filtered X-ray map (which has a FWHM angular resolution of 3 degrees) was correlated with the NVSS map, the
CCF(0) is attenuated by only 1.5\% with respect to the unfiltered CCF(0), again a neglible amount.  So we 
conclude that the high pass filtering has a negligible effect on the above results.

\subsection{Comparison with Other Measurements and Predictions}

There is evidence of excess, small angular scale, fluctuations in the three year WMAP Q band data
\citep{huf06,hin07}.  Under their assumption of uncorrelated (i.e., Poisson only) fluctuations,
it is straightforward to translate this excess to equivalent Q-band Rayleigh-Jeans temperature 
fluctuations in a one square degree beam.  The Hinshaw {\it et al.} (2007) estimate implies {\it rms}
fluctuations of T$_{RJ} \sim 4.5 \mu $K while the Huffenberger {\it et al.} (2006) estimate 
corresponds to T$_{RJ} \sim 4.0 \mu $K.  That these two estimates are consistent with our
rough estimate above should be tempered by the assumptions required to arrive at the latter.
To truly unravel the excess, small scale fluctuations, a sensitive, blind radio survey at the 
high frequencies normally used in CMB studies would be very helpful.  In the meantime, 
our results suggest no unpleasant surprises in the WMAP frequency range.

CMB measurements at smaller angular scales are more prone to 
contamination by foreground sources, since the power spectrum,
$\ell (\ell +1) C_{\ell}$, of 
unclustered sources is proportional to $\ell^2$.  There is a substantial 
literature on the effect of radio sources on the microwave  power 
spectra reported by the CBI, DASI and ACBAR groups, for instance. 
This is nicely summarized by Toffolatti {\it et al.} (2005).  Toffolatti 
and his colleagues also attempt to model the expected contributions 
to high $\ell$ fluctuations from various classes of foreground sources. 
Unfortunately, these predictions can not be directly compared to our
measurement because the models, like the microwave measurements, 
exclude radio sources down to a limit well below the $\sim 1$ Jy cut-off we 
employed.  For instance, the CBI group (Readhead {\it et al.} 2004; Mason 
et al, 2003) removed all sources down to $\sim 30$ mJy (and many weaker sources as 
well) from their images before calculating the power spectrum.  
Therefore the measured (Readhead {\it et al.} 2004) and predicted 
(Toffolatti {\it et al.} 2005) levels of fluctuation lie roughly a factor 
ten lower than our value projected to $\ell = 1000$ and a frequency of 30 GHz.

On the other hand, we can compare our value of $\sim 5~(\nu/40.7$ GHz)$^{-2.1} \mu$K
to values of {\it rms} fluctuation predicted by Toffolatti 
{\it et al.} (1998), since these predictions exclude only sources below 1 
Jy.  Figure 6 of that paper shows that predicted level of 
fluctuations on a $\sim 1$ degree scale is $\sim 4.5 \mu$K.  When we scale from 44 
GHz to 40.7 GHz using a $\beta = -2.1$ temperature spectrum, and correct from 
the Gaussian angular scale Toffolatti used to a 1 degree top hat beam, we obtain 
$\sim 5.3 \mu$K, remarkably close to our estimate.  Further, if we compare temperature 
fluctuations shown in his figures 5 and 6 for 30, 44, 70 and 100 GHz, we find that the 
implied spectral index on scales of $\sim 1$ degree is $\beta \sim -1.9$, close to our 
best-fit value, and inconsistent with a normal synchrotron spectrum.
We conclude that we have found no evidence in the WMAP 
frequency range of a large contribution to the microwave sky not 
captured in Toffolatti's 1998 models.

In the decade since 1998, refined models of source counts at 
high microwave frequencies have been published by, among others, 
Cleary {\it et al.} (2005), De Zotti {\it et al.} (2005), and Waldram {\it et  al.} (2007).  
Unfortunately, these authors do not provide calculations of the expected fluctuation level 
in the microwave sky.  We can, however, compare their source counts 
with those of Toffolatti {\it et al.} (1998) at the appropriate frequency 
and in the relevant range of flux density.   In general, the more 
recent source counts fall roughly a factor $1.5 - 2.0$ below Toffolatti's. 
Thus the expected level of fluctuations would be $20-40\%$ lower than 
Toffolatti's 1998 value, but still in agreement with our 
results.  Obviously, the roughness of our estimate does not allow us 
to discriminate between these various models.  What we can say is 
that, up to 90 GHz at least, there are no surprises in the form of 
unexpected populations of high frequency sources not present in lower 
frequency radio catalogs.

\acknowledgments

We thank K. Jahoda for the original HEAO I A2 data, J. Pober who initiated an
early analysis of residual microwave fluctuations, and J. Peebles for helpful discussions.
We acknowledge the use of the Legacy Archive 
for Microwave Background Data Analysis (LAMBDA). Support for LAMBDA is provided by the 
NASA Office of Space Science.  This work was supported in part by NSF 
grant AST 0606975 to Haverford College.

\end{document}